# Anomalous local magnetic field distribution and strong pinning in CaFe$_{1.94}$Co$_{0.06}$As$_2$ single crystals

Pabitra Mandal[1], Gorky Shaw[1], S. S. Banerjee[1,(a)], Neeraj Kumar[2], S. K. Dhar[2], A. Thamizhavel[2]

[1] Department of Physics, Indian Institute of Technology, Kanpur-208016, India
[2] Department of Condensed Matter Physics and Materials Science, Tata Institute of Fundamental Research, Mumbai- 400005, India

**Abstract** – Magneto-optical imaging of a single crystal of CaFe$_{1.94}$Co$_{0.06}$As$_2$, shows anomalous remnant magnetization within Meissner like regions of the superconductor. The unconventional shape of the local magnetization hysteresis loop suggests admixture of superconducting and magnetic fractions governing the response. Near the superconducting transition temperature, local magnetic field exceeds the applied field resulting in a diamagnetic to positive magnetization transformation. The observed anomalies in the local magnetic field distribution are accompanied with enhanced bulk pinning in the CaFe$_{1.94}$Co$_{0.06}$As$_2$ single crystals. We propose our results suggest a coexistence of superconductivity and magnetic correlations.

**Introduction.** – Superconductivity with conventional singlet pairing is usually considered to be incompatible with magnetic ordering [1]. However unconventional pairing mechanisms could lead to the coexistence of two mutually antagonistic phenomena of magnetism and superconductivity. It is increasingly being recognized that in the iron - pnictide compounds magnetic fluctuations mediate quasi particle pairing [2] resulting in an unconventional isotropic s-wave order parameter which changes sign on different sheets of the Fermi surface and has nodes [2,3]. Doping the iron - pnictide compounds plays an important role in controlling magnetic fluctuations and the onset of superconductivity. Experimentally one finds long range antiferromagnetic (AFM) correlation in the parent pnictide compounds [4]. Subsequent doping leads to the loss of long range magnetic correlations and appearance of superconductivity. Doping dependence studies reveal different scenarios: in some compounds there is an abrupt [5] suppression of AFM order at the onset of superconductivity while in others superconductivity coexists with short range magnetic order [6,7] through microscopic phase separation into pure magnetic and superconducting fractions [7]. Calculations on doping dependent studies in an LaFeAs(OF) iron arsenide system [8] indicate the presence of magnetic fluctuations which could mediate superconductivity [3]. Recent bulk transport and magnetization studies indicate for 3 - 4% Co doping in CaFe$_{2-x}$Co$_x$As$_2$ [9] long range AFM order at ~ 170 K is suppressed followed by the appearance of bulk superconductivity with $T_c(0)$'s in the range of 17 - 20 K.

Recent μSR studies on the distribution of magnetic field inside SrFe$_{1.75}$Co$_{0.25}$As$_2$ suggest the external magnetic field induces magnetic ordering [10] in the superconducting state of this system. These studies have motivated suggestions that magnetic ordering may act as an effective mechanism for enhanced pinning in pnictide superconductors [11]. In this letter, we image the local magnetic field distribution at low fields in a single crystal of CaFe$_{1.94}$Co$_{0.06}$As$_2$ using high-sensitivity Magneto-optical imaging (MOI) technique [12,13,14] (cf. section I of supplementary information in ref. [14] for additional details on the MOI technique employed). We find anomalous weak remnant magnetization within Meissner like regions of the superconductor. Isofield temperature dependent MOI studies reveal that the diamagnetic superconducting regions transform into regions with positive magnetization close to $T_c(0)$. By comparing with bulk magnetization data we find evidence for strong pinning in our samples. Our results suggest that near $T_c$ the weak diamagnetic response of the superconductor uncovers the presence of magnetic correlations in the sample.

**Experimental details.** – For our study we use single crystals (cf. Neeraj Kumar et al., ref. [9]) of CaFe$_{1.94}$Co$_{0.06}$As$_2$, with dimensions 0.66 × 0.47 × 0.064 mm$^3$ (bulk superconducting transition temperature, $T_c(0)$ ~ 17 K determined from bulk magnetization and transport measurement). Results are verified on two single crystals with the same stoichiometry. Details of the growth of the single crystals, their quality, x-ray diffraction measurements confirming the tetragonal structure and bulk characterization of the sample via specific heat, transport and dc magnetization measurements of the single crystals have been presented elsewhere (Neeraj Kumar *et al.*, ref. [9]). Spatial homogeneity of stoichiometry in the single crystals on local and macroscopic length scales has been verified via EPMA and spot EDX given in section II of supplementary information at ref. [14]. We use high-sensitivity MOI technique [13,14] for imaging the local magnetic field distribution across the sample surface, *viz.*, $B_z(x,y)$ ($z$-component is perpendicular to the ab-plane (surface) of the single crystal), and the external - magnetic field ($H$) is applied parallel to the crystallographic c-axis. Prior to performing MOI, the sample was freshly cleaved to obtain a flat surface. Note that due to relatively long acquisition times ($\sim$ few tens of seconds) the present MOI technique is not sensitive to time-varying magnetization response.

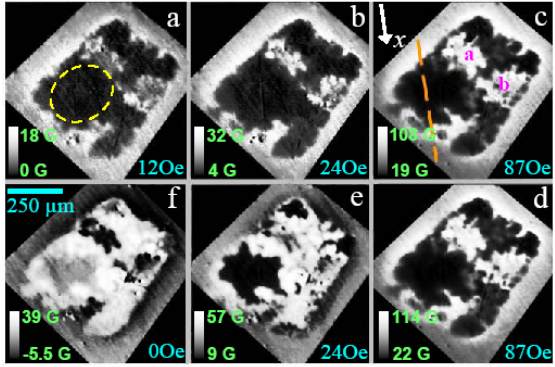

**Fig. 1:** MO images at (a) 12 Oe, (b) 24 Oe and (c) 87 Oe at $T$ = 11K, on increasing $H$ (virgin run). Lower panel shows images at the same field values while decreasing $H$ from 90 Oe (reverse run), except fig. 1(f) corresponds to $H = 0$ Oe. The gray scale bars beside each image indicate the minimum and maximum values of local fields encountered for each MO image.

**Local magnetic field distribution measured as a function of changing external field.** – Figure 1 shows magneto optical (MO) images captured at 11 K. Figures 1(a) to 1(c) represent images obtained while increasing $H$ (virgin run), while figs. 1(d) to 1(f) are captured while reducing $H$ (reverse run) from 90 Oe (image at 90 Oe not shown). In fig. 1(a) the regions with dark contrast have a *mean* local field (within the yellow dashed) encircled region), $\langle B_z \rangle \sim 0$ G which are the Meissner-like regions of the superconductor, (note, the mean $B_z$ is denoted by $\langle B_z \rangle$ where the mean is computed by averaging $B_z(x,y)$ distribution over an area of just $\sim$100 μm$^2$ region (10 μm $\times$ 10 μm) in the sample). The portions of the image with grayish contrast (e.g. just above the yellow dashed circle in fig. 1(a)) possess nonzero$\langle B_z \rangle$, due to vortex penetration in these regions of the sample. With increasing $H$ the grayish regions turn brighter. In fig. 1(c) over the bright regions labeled a and b, the mean $\langle B_z \rangle$ increases to 97 G which is greater than $H = 87$ Oe, indicating a region where anomalous positive magnetization feature, $\langle B_z \rangle > H$, has developed (we shall discuss this in detail in subsequent section). Regions with grayish or dark contrast persist with a diamagnetic response $\langle B_z \rangle < H$. Up to $H = 90$ Oe the presence of dark regions (like the region encircled by dashed circle in fig.1(a)) in the MO image suggests the persistence of Meissner like regions which shield the external $H$ up to 90 Oe in the sample. Thus locally $H = 90$ Oe is less than the lower critical (or penetration) field of the superconductor. In the images of the lower panel (figs. 1(d)-1(f), as $H$ is reduced from 90 Oe, the Meissner like shielded regions (dark contrast) instead of maintaining their dark contrast down to 0 Oe, appear to progressively become brighter. Note in fig. 1(f) that at 0 Oe on the reverse run, there is a significant brightening indicative of remnant flux present within the Meissner like region when $H$ is reversed from 90 Oe, a field value which is below the *local* lower critical field for the dark region. A Meissner phase being a perfectly diamagnetic state with zero flux cannot possess any remnant flux when $H$ is cycled back to zero from below the lower critical field of that region. (The line-like light-gray features running diagonally across the image in fig. 1(f) are defects on the magneto-optical indicator placed on the sample for imaging.)

The unusual observation on fig.1 is investigated and confirmed further by calculating the local field distribution ($B_z(x)$) along the dashed line (drawn in fig. 1(c)) in figs. 2(a) and 2(b) at different $H$ at 11 K. We define $x$-coordinate to be along the line (cf. fig. 1(c)). The dashed vertical lines in figs. 2(a) and 2(b) identify the sample edges encountered along the line. The enhanced $B_z(x)$ near the sample edges is associated with geometric barrier-related shielding of $H$ from sample interior [15]. With increasing $H$, fig. 2(a) shows a Bean-like [16] gradient in $B_z$ indicating penetration of vortices into the sample interior from the edges. Due to the large thickness to width ratio of the crystals $\sim$ 0.13, the gradients appear straight [17] and from the slope of the linear portion of the $B_z(x)$ profile in fig.2(a), we estimate the critical current density $J_c(90$ Oe, 11K$) \sim dB_z/dx \sim 7 \times 10^3$ A/cm$^2$. The estimated $J_c$ is comparable to the typical value reported in literature on single crystals of other pnictides [18].

The periphery of the Meissner-like region identified in fig.1(a) (within the dashed circle) is identified in fig. 2(a) as the region AB located between the two arrows, where $B_z \sim$ constant inside AB with almost zero slope. From fig. 2(a) it appears that the region of the sample within AB maintains a Meissner like response with no gradient in $B_z$ up to fields of 90 Oe. Closer inspection, however, shows that within AB region there is an anomalous increase in $\langle B_z \rangle$ from 0 G up to 20 G as $H$ is increased, which we argue is intrinsic and not an experimental artifact. This observation is inconsistent with the region between AB being a perfectly shielded Meissner state. We have confirmed this anomalous behavior at other $T$'s and in other samples of the same stoichiometry. In fig. 2(b), on decreasing $H$ from 90 Oe to 0 Oe, conventional reversed Bean like gradients near the sample edges are observed. From the

$B_z(x)$-profile in fig. 2(b) note that $B_z$ within AB region does not decrease to zero but remains at a finite remnant value of $\sim$ 3.5 G. It is this remnant $B_z \sim 3.5$ G which is responsible for the bright contrast in the Meissner like region of the sample at $H$ = 0 Oe as noted for fig. 1(f). Thus within the Meissner region of the sample it appears that the diamagnetic $((B_z - H) < 0)$ gradientless $B_z(x)$ response within AB is not zero and it rides on a uniform background $B_z$ which increases with $H$ and is hysteretic.

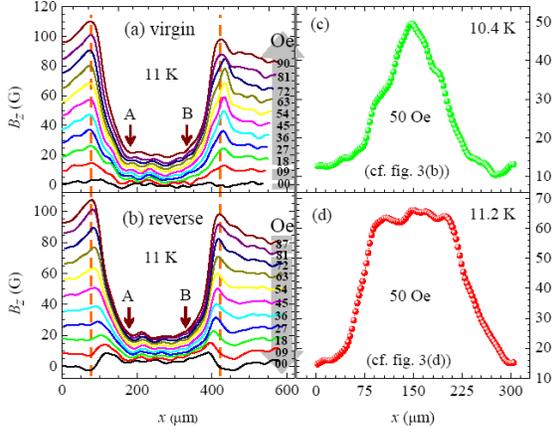

Fig. 2: $B_z$ vs. $x$ profile along the line shown in fig. 1(c), at different $H$ for (a) virgin and (b) reverse run. Two vertical dashed lines represent sample edges. (c), (d) $B_z$-$x$ profiles at 10.4 K and 11.2K determined from figs. 3(b), 3(d) respectively.

Note that this behavior is observed only for the magneto-optical response over the sample and not outside the sample. It appears that within this Meissner like region there exists an unexpected response. To ascertain that the observations presented above is not unique to one sample, not related to any peculiar realization of inhomogeneity in the chemical composition of this sample, we have confirmed the presence of a similar features within a gradientless Meissner like region in another single crystal of CaFe$_{1.94}$Co$_{0.06}$As$_2$ obtained from the same batch (cf. section III in supplementary information in Ref. [14] and compare the behavior in the region between the arrows with that in figs.2(a) and 2(b)).

**Local magnetic field distribution measured as a function of temperature at fixed field, transformation to positive magnetization.** – Shown in figs. 3(a)-(e) are representative MO images captured at stabilized different $T$ in an $H$ = 50 Oe ($H \parallel c$, with zero field cooled (ZFC) history, sample is cooled to $T < T_c$ ($\sim$17K) before applying $H$). At 10 K (fig. 3(a)), we observe regions marked 4 and 5 appear significantly brighter than the rest of the sample. In these regions $\langle B_z \rangle \sim 75$ G > $H$ (= 50 Oe). Similar bright regions with $\langle B_z \rangle > H$ at the same locations are seen in fig. 1 (see regions marked a, b in fig. 1(c)). Other regions in fig. 3(a) are comparatively darker (see grayish regions 1, 2 and 3) with

$\langle B_z \rangle < H$, for e.g., in region 2, $\langle B_z \rangle \sim 10$ G (< 50 Oe) at 10.4 K. In the light gray region marked 1 in fig. 3(a), vortices have already penetrated the superconductor. As $T$ is increased (cf. figs. 3(a), 3(b) and 3(c)), it is clear that regions like 1 with grayish contrast not only expand in size due to the penetrating vortices but progressively become brighter (cf. figs. 3(c) and 3(d)). The $B_z(x)$ distribution across the line in fig. 3(b) is shown in fig. 2(c).

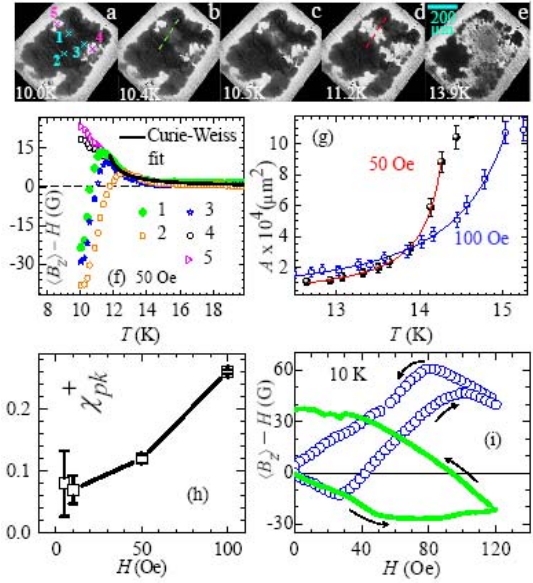

Fig. 3: MO images for 50 Oe ($H \parallel c$, ZFC) at (a) 10.0 K, (b) 10.4 K, (c) 10.5 K, (d) 11.2 K and (e) 13.9 K. (f) local ($\langle B_z \rangle$-$H$) vs. $T$ calculated at different regions indicated in panel (a). Solid line shows a Curie-Weiss fit to the curves. Compare the positive peak like feature in ($\langle B_z \rangle$-$H$) with a similar feature seen in fig. 4(a) inset. (g) $T$ dependence of the area ($A$) with positive ($\langle B_z \rangle$-$H$). Solid curves show the fitting to the equation, $A(T) \propto 1/|T\text{-}T_{mgn}|^\gamma$. (h) $\chi_{pk}^+(H)$ vs. $H$ plot (cf. text for detail). (i) Open blue circle shows local (for region b marked in fig. 1(c)) and solid green curve shows bulk ($\langle B_z \rangle$-$H$) vs. $H$ behavior at 10 K (cf. text for details).

The $B_z(x)$ shows a maximum $\sim$ 50 G = $H$ at 10.4 K. The $B_z(x)$ in fig. 2(d) shows that across the line in fig. 3(d), the bright region (of width $\sim$ 150 $\mu$m) possesses a $B_z \sim 65$ G > $H$ = 50 Oe implying positive ($B_z - H$). Note that this region of $\sim$ 150 $\mu$m width having significantly positive magnetization has a uniform field distribution devoid of any critical state like gradients in $B_z(x)$. This suggests that the anomalous positive magnetization feature we observe is not related to any pinning related inhomogeneities in the magnetic flux distribution. In any particular region of the sample we find that as $T$ increases, $B_z$ increases to reach $B_z = H$ and further continues to increase beyond $H$ reaching a maximum $B_z$ value after which it decreases to zero near $T_c(H)$. This behavior is summarized in fig. 3(f), where ($\langle B_z \rangle$-$H$) is calculated from the MO images and plotted as a function of $T$ for five different regions marked in fig. 3(a). We argue that it is unlikely that chemical inhomogeneities in the sample would produce the kind of spreading front like behavior with increasing $T$ of regions where negative ($\langle B_z \rangle$-$H$) transforms to a positive ($\langle B_z \rangle$-$H$) as $T$ increases towards $T_c$ (cf. fig.3 and movie at section IV of supplementary information at ref. [14]).

**Transformation from a diamagnetic to paramagnetic response as $T$ approaches $T_c$.** – A notable feature in ($\langle B_z \rangle$-$H$) *vs.* $T$ plot of fig. 3(f) is that at low $T$ (= 10 K), region 4 and 5 do not exhibit a diamagnetic response which is characteristic of a ZFC superconductor. Instead, in these regions (4 and 5) we observe that local magnetization, ($\langle B_z \rangle$-$H$), is significantly positive at low $T$ and its value decreases as $T$ is increased toward 14 K. This feature of positive magnetization is neither characteristic of a sample which has been ZFC, nor is it characteristic of the near zero or a very weakly diamagnetic field-cooled response [19,20] expected of a conventional superconductor with pinning. The weak diamagnetic response is expected not only on the average magnetic flux distribution is uniform over the bulk of the sample but is also expected in a local magnetization measurement when on the scale of the local measurement, the course grain average (region over which averaging is performed spans over a collection of vortices) field distribution is uniform. It is tempting to associate the nature of the local ($\langle B_z \rangle$-$H$)-($T$) curves in fig.3(f) observed for regions 4 and 5 with magnetic ordering. In the vicinity of 10 K, fig. 3(f) shows that regions 1, 2 and 3 possess a response typical of a ZFC diamagnetic superconductor, having $\langle B_z \rangle$ < $H$. With increasing $T$, the magnetization approaches zero due to a decrease in diamagnetic shielding currents flowing in the superconductor. Above 11 K, the magnetization becomes positive and goes through a peak positive value before decreasing monotonically towards zero as $T$ approaches $T_c$. In fig. 3(h) we plot $\chi^{+}_{pk}(H) = ((\langle B_z \rangle - H)^{+}_{pk})/H$ where ($\langle B_z \rangle - H)^{+}_{pk}$ is the peak value of the positive magnetization ($\langle B_z \rangle - H$), denoted as ($\langle B_z \rangle - H)^{+}$. We observe an increasing trend of peak $\chi^{+}_{pk}$ with $H$, which is opposite to the behavior seen for conventional and high $T_c$ superconductors associated with the paramagnetic Meissner effect (PME) [21] where a similar positive magnetization peak near $T_c$ is observed for superconductors field cooled in a small applied field. In conventional PME, $\chi^{+}_{pk}(H)$ decreases with increasing $H$ [21]. As argued below, the unusual behavior of $\chi^{+}_{pk}(H)$, we suggest, is associated with magnetic correlations in the system. The positive nature of ($\langle B_z \rangle$-$H$) is uncovered only when the diamagnetic shielding response of the superconductor weakens close to $T_c$. The increasing behavior of $\chi^{+}_{pk}$ with $H$ suggests that the ($\langle B_z \rangle - H)_{pk}$ is increasing faster than $H$. A similar feature of $\langle B_z \rangle$ > $H$ was noted for fixed $T$ varying $H$ measurements (cf. fig. 1(c) discussion).

Similar to our results, Ref.[10] also suggests the presence of magnetic correlations as indicated by an enhancement in the local magnetic field with decreasing $T$, in the superconducting state of a pnictide superconductor.

**Anomalies in the local magnetization hysteresis loop and the onset of magnetic correlations.** – Figure 3(i) represents the local (open blue circles) and bulk (solid green curve) magnetization, ($\langle B_z \rangle$-$H$) *vs.* $H$, response at 10 K calculated from the MO images using [22] ($\langle B_z \rangle$-$H$) = $(\int_A [B_z(x,y)-H]dxdy)/(\int_A dxdy)$, where $A$ is integration area ($A$ is the full sample area for solid green curve and for open blue symbol, $A$ is a $10 \times 10$ μm² region located within the white region b marked in fig.1(c) and which is identified with region 3 in fig.3(a) possessing positive ($\langle B_z \rangle$-$H$) locally). Notice that the solid green curve has a shape similar to a conventional superconducting hysteresis loop and similar bulk loops have been reported for pnictides [18]. The shape of the local hysteresis loop (open blue circles) is unlike that of a conventional superconductor. In fact a first glance at the peculiar shape of the local ($\langle B_z \rangle$-$H$)-($H$) curve suggests similarities with a ferromagnetic hysteresis loop. The diamagnetic linear variation of ($\langle B_z \rangle$-$H$)-($H$) curves till 30 Oe for both the bulk and local behavior confirms the Meissner like shielding response (a response of the type seen within the dashed (yellow circled) region in fig.1(a)). While the bulk ($\langle B_z \rangle$-$H$)-($H$) continues its linear behavior up to a larger $H$ = 50 Oe (due to geometric barriers affecting the penetration field), the local ($\langle B_z \rangle$-$H$)($H$) turns around beyond 30 Oe and crosses over to a positive magnetization (($\langle B_z \rangle$-$H$)>0) regime beyond 45 Oe. As the field is reversed from 120 Oe, the local ($\langle B_z \rangle$-$H$)($H$) loop reveals a remnant magnetization of ~ 3 to 4 G (10 K). The unusual shape of the local hysteresis loop discussed above suggests the possibility of magnetic correlations coexisting along with superconductivity.

Figure 3(g) shows the increase in area ($A$) of the region with positive ($\langle B_z \rangle$-$H$) response with $T$ at 50 Oe and 100 Oe (cf. fig. 3(a)-(e) and movie link at section IV of supplementary information at ref. [14]). We observe that the area with positive magnetization grows and spreads across the sample with increasing $T$ thus suggesting suggests its origin is not associated with sample inhomogeneity. The behavior of $A(T)$ fits to $A(T)$ α $1/|T-T_{mgn}|^{\gamma}$, where at 50 Oe the critical exponent $\gamma$ = 1.12 ± 0.10 and the mean temperature scale $T_{mgn}$ =14.54 ± 0.06 K and at 100 Oe $\gamma$ = 1.59 ± 0.12 and $T_{mgn}$ =16.06 ± 0.14 K. The above divergence in area suggests a diverging correlation length. The correlation length below $T_{mgn}$ is associated with the onset of magnetic correlations in the sample producing a positive magnetization response.

The $T_{mgn}$ is a critical temperature associated with the onset of enhanced magnetic correlations found in close proximity to $T_c$. The above inference is supported by the observation of a Curie-Weiss fit to ($\langle B_z \rangle$-$H$) *vs.* $T$ response (cf. solid black curve in fig. 3(f)). As one reduces $T$ below $T_c$, enhanced magnetic correlations lead to a positive feature in the local ($\langle B_z \rangle$-$H$)-($T$) response. However with further lowering of $T$ the region over which the correlations exist shrinks at the expense of a growing superconducting fraction with a predominating diamagnetic response, leading to a positive peak like feature in $(\langle B_z \rangle - H)$-($T$). At low $T$ the area over which magnetic correlations survive may fall below the resolution of our setup. Evidence for magnetic order coexisting with superconductivity at low $T$ is suggested by the monotonic variation of $\chi^+_{pk}(H)$ (cf. fig. 3(h)) from the microscopic magnetically ordered regions which produce a uniform and hysteretic offset in $B_z(x)$ response between Meissner regions as discussed earlier for figs.2(a) and 2(b). The Curie Weiss fit $\chi = c/(T-T_{Curie})$ to the positive decaying portion of ($\langle B_z \rangle$ - $H$)-($T$) curve at 50 Oe (cf. solid black curve in fig. 3(f)) yields a mean $T_{Curie} \simeq 12$ K (the Curie $T$), and $c = N\mu_{eff}^2 = 1.207$ cc K/mole, where $N$ is the molar density of magnetic moments with effective magnetic moment $\mu_{eff}$. The low value of $T_{Curie}$ suggests that magnetic exchange interactions are weak. By using $N$ = Avogadro number, we estimate the magnetic moment of $\sim 1.5 \mu_B$/(Fe-atom) ), which is close to the theoretically proposed Fe moment value [23] in CaFe$_2$As$_2$ parent pnictide compound. Alternatively, if we assume that not all Fe atoms in the material contribute to the moment, therefore using a reported value [24] of Fe $\mu_{eff} = 0.75(4)\mu_B$ in $c = N\mu_{eff}^2$, we estimate $N \sim 25\%$ of Avogadro number, which implies a mean moments spacing of $\sim 2(d_{Fe-Fe}$ spacing = 2.7 Å). Note, Co concentration of 3% in CaFe$_{1.94}$Co$_{0.06}$As$_2$ corresponds to an average Co-Co spacing of $\sim 9$ Å. Recent STM study on the FeSe system [2] suggests that it has superconducting state with s$^\pm$ order parameter with nodes associated with strong nearest neighbour (NN) exchange interactions between Fe sites. Our study suggests that in CaFe$_{1.94}$Co$_{0.06}$As$_2$, exchange interaction between next to nearest neighbor (NNN) Fe sites maybe playing an important role in mediating weak magnetic correlations to coexist along with superconductivity. Figure 4(c) shows the temperature dependence of the lower critical field $H_{c1}$ (demagnetization corrected) which is estimated using deviation from Meissner linear behavior in the isothermal bulk magnetization $M$ vs. $H$ measurements as shown by an upward arrow in the bulk $M$-$H$ loop of 14K in fig. 4(b). Down to reduced temperatures $t = (T/T_c(0)) \sim 0.6$, with $T_c(0) \sim 17.8$ K (determined from extrapolation of the $H_{c1}(T)$ data), the $H_{c1}(t)$ behaves as (1-$t$) (cf. fig. 4(c)), which is unlike a BCS superconductor.

**Bulk magnetization and strong pinning regime close to $T_c$.** –Shown in fig. 4(a) is $4\pi M$ vs. $T$ data measured on the Quantum design SQUID magnetometer ($H \parallel c$). The sample was cooled from 30 K in $H = 0$ Oe down to 5 K and the isofield $4\pi M$ vs. $T$ was recorded in an $H = 50$ Oe while warming up the sample. Note that at 5 K, the sample exhibits a strong diamagnetic shielding response indicating bulk superconductivity sets in the sample at low $T$. As $T$ approaches 17 K ($\sim T_c(0)$), the diamagnetic response decreases. In the inset of fig. 4(a), $M$ instead of monotonically approaching zero becomes positive and goes through a peak before reducing to zero. This feature is similar to that observed in local measurements in fig.3(f). We also find unusual features in specific heat behavior in the regime near $T_c$ where we find evidence for magnetic fluctuations in our local measurements (cf. supplementary information, section V, at Ref. [14]).

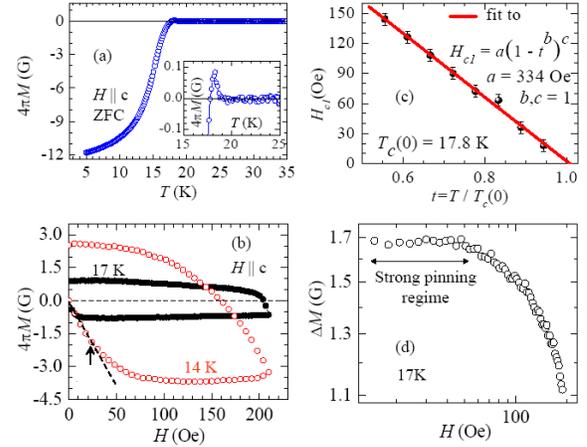

**Fig. 4:** (a) Bulk $M$ vs. $T$ curve (ZFC warming, $H$ ($\parallel$c) = 50 Oe). The inset shows positive peak near $T_c$ (see text for details). (b) Bulk $M$-$H$ hysteresis loops measured in the SQUID magnetometer at 14K and 17K. The arrow denotes the location from where $H_{c1}$(14K) is determined. (c) $H_{c1}$ vs. $t$ (= ($T/T_c(0)$) behavior calculated from bulk $M$-$H$ measurements. The solid line shows the data fitted to the form $H_{c1} = a(1-t^b)^c$ with $a$ = 334 Oe, $b$, $c$ = 1 giving the best fit. (d) $\Delta M$ (17K) vs. $H$ on log-log scale for $H > 20$ Oe.

**Evidence of strong pinning near $T_c$ via bulk magnetization measurements.** – In recent times origin of strong pinning in pnictides has been attributed to being either chemical in nature, related to fluctuations in doping concentration on the scale of few 100 nm [25] or due to magnetic fields related to onset of magnetic ordering in the pnictides [10,11]. Our data suggests magnetic correlations as being responsible for pinning in the Co doped CaFe$_2$As$_2$ compound. To investigate the pinning in our sample we capture bulk $M$-$H$ hysteresis loops measured in the SQUID magnetometer above 10 K. Shown in fig.4(b) are the virgin and reverse runs of the $4\pi M$ vs. $H$ hysteresis loop (upto low $H = 210$ Oe) measured at $T = 14$ K and 17 K. Here we find an unusual feature, i.e., at 17 K which is $\sim 0.95$ $T_c$ (with $T_c(0) \sim 17.8$ K determined from fig.4(c)), significant amount of hysteresis persists. Figure 4(d) shows the behavior of the width of hysteresis loop ($\Delta M$) as a function of the applied magnetic field. Note that $\Delta M \propto J_c$, is a measure of pinning strength in a superconductor. Strong intervortex interactions in a weak pinning environment typically leads to a weak pinning scenario where $J_c$ (or equivalently $\Delta M$) is proportional to $1/H^\alpha$ [26], where $\alpha$ is a positive constant. However, the log-log plot in fig.4(d) at 17 K shows that $\Delta M$ does not follow the weak pinning behavior, rather up to $\sim 80$ Oe, the $\Delta M$ is almost field independent. The strong pinning regime is expected in the limit of dilute density of vortices wherein the vortices are weakly interacting, i.e., the mean intervortex separation ($a_0$) is greater than the superconducting penetration depth ($\lambda$). In such a situation, the vortices can be easily pinned on the pin-ning sites and the vortex state enters a maximally pinned state. In this strong pinning regime, the $J_c$ or $\Delta M$ is expected to be almost field independent, as changes in $H$ in this regime do not significantly enhance the intervortex interaction to weaken the effective pinning strength.

In fig.4(d) we find the field independent $\Delta M(H)$ regime *i.e.*, the strong pinning regime extending up to 80 Oe and surviving at 17 K ($\sim 0.95$ $T_c$). Near $T_c$, large thermal fluctuations are expected to weaken the pinning potential significantly. In such an $H$ - $T$ regime, the thermal smearing effects should have been significantly weakened the pinning. Furthermore, at 80 Oe where the intervortex separation is $a_0 \sim 400$ nm, due to the divergence of $\lambda(T)$ at $T = 17$ K $\sim 0.95T_c$, it is expected that the vortices aren't weakly interacting. Therefore near 17 K $\sim 0.95T_c$ one does not expect to find a strong pinning regime. However our observation shown in fig.4(d) suggests strong pinning surviving at $T$ close to $T_c$ which suggests the presence of strong pinning centers in the superconductor whose pinning potential is much larger than the thermal ($\sim k_B T_c$) energy scale.

In fig.3 we have already argued that the positive $\langle \langle B_z \rangle - H \rangle$ response near $T_c$ at low fields is indicative of the presence of magnetic correlations in the superconductor. Regions with magnetic correlations coexisting along with superconductivity and having ordering temperature higher than $T_c$ of the superconductor would be suitable candidates whose pinning strengths would be effective up to close to $T_c$. Therefore, in the present pnictide system investigated, we suggest magnetic correlations as a plausible source of strong pinning surviving in the superconductor up to $\sim T_c$.

To summarize, at low $T$ ($<< T_c$) while the material exhibits features like any bulk superconductor, with increasing $T$ as the shielding response of the superconductor weakens, the presence of underlying magnetic correlations in the $CaFe_{1.94}Co_{0.06}As_2$ is uncovered. We suggest that $CaFe_{1.94}Co_{0.06}As_2$ is a rich system to investigate aspects related to the coexistence of magnetic correlations along with superconductivity.

SSB acknowledges DST, CSIR and MHRD-India, and Debanjan Chowdhury for his help during the experiments.

## REFERENCES

[1] Ginzburg V. L., Sov. Phys. JETP **4** (1957) 153; Saint James D., Sarma G. and Thomas E. J., Type II Superconductivity (Pergamon, New York, 1969).

[2] Can-Li Song *et al.*, *Science*, **332** (2011) 1410.

[3] Mazin I. I. *et al.*, *Phys. Rev. Lett.*, **101** (2008) 057003; Kazuhiko Kuroki *et al.*, *Phys. Rev. Lett.*, **101** (2008) 087004.

[4] De la Cruz C. *et al.*, *Nature* (London) **453** (2008) 899 ; Zhao J. *et al.*, *Nature Mater.*, **7** (2008) 953.

[5] Kamihara Y., Watanabe T., Hirano M., and Hosono H., *J. Am.Chem. Soc.*, **130** (2008) 3296.

[6] Drew A. J. *et al.*, *Nat. Mater.*, **8** (2009) 310; Bernhard C. *et al.*, *New Journal of Physics*, **11** (2009) 055050.

[7] Aczel A. A. *et al.*, *Phys. Rev. B*, **78** (2008) 214503; Park J. T. *et al.*, *Phys. Rev. Lett.*, **102** (2009) 117006.

[8] Singh D. J. and Du M. H., *Phys. Rev. Lett.*, **100** (2008) 237003.

[9] Neeraj Kumar *et al.*, *Phys. Rev. B*, **79** (2009) 012504; Marcin Matsiak, Zbigniew Bukowski and Janusz Karpinski, *Phys. Rev. B*, **81** (2010) 020510(R).

[10] Khasanov R. *et al.*, Phys. Rev. Lett. **103** (2009) 067010.

[11] Inosov D. S. *et al.*, *Phys. Rev. B*. **81** (2010) 014513.

[12] Magneto-Optical Imaging, edited by Tom H. Johansen and Daniel V. Shantsev, *Nato sci. ser.* II, vol. 142, Kluwer academic publisher, The Netherland (2004).

[13] Rinke J. Wijngaarden *et al.*, *Review of Scientific Instruments*, **72** (2001) 2661.

[14] For additional details, see different sections of supplementary information file at http://home.iitk.ac.in/~satyajit/CaFeCoAs2/CaFeCoAs2_supp_info.pdf

[15] Dan T. Fuchs *et al.*, *NATURE*, **391** (1998) 373-376; James S.S. *et al.*, *Physica C*, **332** (2000) 173–177; Zeldov E. *et al.*, *Europhys. Lett.*, **30** (1995) 367- 372; Chikumoto N. *et al.*, *Phys. Rev. Lett.*, **69** (1992) 1260.

[16] BEAN C. P., *Phys. Rev. Lett.*, **8** (1962) 250.

[17] Brandt E. H., *Phys. Rev. B*, **54** (1996) 4246.

[18] Prozorov R. *et al.*, *Phys. Rev. B*, **81** (2010) 094509.

[19] Prozorov R. *et al.*, *Phys. Rev. B*, **82**, (2010) 180513 (R).

[20] Yamamoto A. *et al.*, *Supercond. Sci. Technol.*, **21** (2008) 095008.

[21] Sigrist M., Rice T.M., *Rev. Mod. Phys.* **67** (1995) 503; Koshelev A. E. and Larkin A. I., *Phys. Rev. B* **52**(1995) 13559; Pradip Das *et al.*, Phys. Rev. B **78** (2008) 214504.

[22] Bartolome E. *et al.*, *Phys. Rev. B*, **72** (2005) 024523; Majer D. *et al.*, *Phys. Rev. Lett.*, **75**, (1995) 1166.

[23] T. Yildirim, *Phys. Rev. Lett.* **102**, 037003 (2009)

[24] Prokes K. *et al.*, *Phys. Rev. B*, **83** (2011) 104414.

[25] Van der Beek C J *et al.*, *Phys. Rev. B* **81** (2010) 174517.

[26] A. I. Larkin and Y. N. Ovchinnikov, J. Low Temp. Phys. **34** (1979) 409; G. Blatter *et al.*, *Rev. Mod. Phys.* **66** (1994) 1125; Shyam Mohan *et al.*, *Phys. Rev. Lett.* **98** (2007) 027003.